\documentclass[useAMS,usenatbib]{mn2e}
\usepackage{graphicx}

\title[On the incidence of magnetic fields in slowly-pulsating B, $\beta$ Cephei and B-type emission line stars]{On the incidence of magnetic fields in slowly-pulsating B, \\ $\beta$ Cephei and B-type emission line stars}
\author[J. Silvester, C. Neiner, H.F. Henrichs, G.A. Wade, et al.]
{J. Silvester$^{1,2}$, C. Neiner$^{3}$, H.F. Henrichs$^{4}$, G.A. Wade$^{2}$, V. Petit$^{5}$, E. Alecian$^{2}$, 
\newauthor {A.-L. Huat$^{6}$, C. Martayan$^{6,7}$, J. Power$^{2}$, O. Thizy$^{8}$}\\
$^{1}$Department of Physics, Engineering Physics \& Astronomy, Queen's University, Kingston, Ontario, Canada, K7L 3N6\\
$^{2}$Department of Physics, Royal Military College of Canada, P.O. Box 17000, Station `Forces', Kingston, Ontario, Canada, K7K 7B4\\
$^{3}$GEPI, Observatoire de Paris, CNRS, place Jules Janssen, 92190 Meudon Cedex, France\\
$^{4}$Astronomical Institute "Anton Pannekoek", University of Amsterdam, Science Park 904, 1098 XH Amsterdam, Netherlands\\
$^{5}$D\'epartement de physique, g\'enie physique et optique, CRAQ, Universit\'{e} Laval, Qu\'{e}bec, Canada, G1K 7P4 \\
$^{6}$GEPI, Observatoire de Paris, CNRS, Universit\'e Paris Diderot; 5 place Jules Janssen, 92190 Meudon, France \\
$^{7}$Royal Observatory of Belgium, 3 avenue circulaire, 1180 Brussels, Belgium \\
$^{8}$Shelyak Instruments, Les Roussets, 38420 Revel, France }  

\begin{document}

\date{Accepted . Received }

\pagerange{\pageref{firstpage}--\pageref{lastpage}} \pubyear{2007}

\maketitle

\label{firstpage}

\begin{abstract}
We have obtained 40 high-resolution circular spectropolarimetric measurements of 12 slowly-pulsating B (SPB) stars, 8 $\beta$ Cephei stars and two Be stars with the ESPaDOnS and NARVAL spectropolarimeters. The aim of these observations is to evaluate recent claims of a high incidence of magnetic field detections in stars of these types obtained {\bf using low-resolution spectropolarimetry} by \cite{hubrig:2006, hubrig:2007, hubrig:2009}. The precision achieved is generally comparable to or superior to that obtained by Hubrig et al., although our new observations are distinguished by their resolution of metallic and He line profiles, and their consequent sensitivity to magnetic fields of zero net longitudinal component. In the SPB stars we confirm the detection of magnetic field in one star (16 Peg), but find no evidence of the presence of fields in the remaining 11.  In the $\beta$~Cep stars, we detect a field in $\xi^{1}$~CMa,  but not in any of the remaining 7 stars.  Finally, neither of the {\bf two B-type emission line} stars shows any evidence of magnetic field. Based on our results, we conclude that fields are not common in SPB, $\beta$ Cep and {\bf B-type emission line stars}, consistent with the general rarity of fields in the broader population of main sequence B-type stars. {\bf A relatively small, systematic underestimation of the error bars associated with the FORS1 longitudinal field measurements of \cite{hubrig:2006, hubrig:2007, hubrig:2009} could in large part explain the discrepancy between their results and those presented here.}
\end{abstract}


\section{Introduction}

Some B-type stars have been established to host strong, organised magnetic fields (e.g. \citealt{bohlender:1987}, \citealt*{bohlender:1993}). The first such stars discovered were main sequence helium-weak and helium-strong stars (\citealt{borra:1979}, \citealt*{borra:1983}) - objects which also display strong photospheric chemical abundance anomalies. The lack of similar anomalies in spectra of the large majority of other B-type stars was taken as evidence that magnetic fields are relatively rare in such stars. It therefore came as a surprise when \cite{henrichs:2000a} reported the detection of a 300~G dipolar magnetic field in the chemically normal, pulsating B1IV star $\beta$~Cephei. This discovery was soon followed by reports of magnetic fields in the $\beta$~Cep star V2052 Oph \citep[][c]{neiner:2003c}, the slowly-pulsating B (SPB) star $\zeta$~Cas \citep[][a]{neiner:2003a} and the classical Be star $\omega$ Ori \citep[][b]{neiner:2003b}. More recently, \cite{donati:2006} have reported a field in the bright B0.5V star $\tau$~Sco, and \cite{petit:2008} and \cite{alecian:2008a} have established the existence of fields in the very young early B-type stars HD 36982 and HD 37061 (in the ONC), NGC 6611-601 and NGC 2244-201. 

The discovery of fields in this diverse population of B-type stars has generated significant interest in their magnetic properties, for several reasons. These stars exhibit a tremendous range of physical phenomena (pulsation, rapid rotation, mass loss, accretion and decretion discs, etc.) capable of highlighting the interaction of the magnetic field with the stellar plasma and their intense radiation fields. Many also end their lives in type II supernovae \citep{stankov:2005}, and are progenitors of (strongly magnetic) neutron stars.

Recently, \cite{hubrig:2006, hubrig:2009} (hereinafter referred to as H06 and H09 respectively) reported longitudinal magnetic field measurements of 61 pulsating {\bf or candidate pulsating} B-type stars (45 SPB stars and 16 $\beta$~Cep stars), acquired using the FORS1 spectropolarimeter at the ESO VLT. In this large sample, H09 claimed detection (with $>3\sigma$ confidence) of 21 magnetic fields in the 45 SPB stars, corresponding to 47 percent of the SPB sample. They also reported the detection of fields in five of 16 $\beta$~Cep stars. In addition, \cite{hubrig:2007}, hereinafter referred to as H07, performed a similar analysis of observations of 15 classical Be stars, reporting significant detections of fields in three of these targets. These results suggest that magnetic fields may be significantly more common in pulsating B-type stars and Be stars than was previously assumed, and possibly that all SPB stars host organised magnetic fields. 

To verify these potentially very important results, we have undertaken high-resolution circular polarisation spectroscopy of a sample of 22 SPB, $\beta$~Cep and Be stars using the powerful ESPaDOnS and NARVAL high-resolution spectropolarimeters at the Canada-France-Hawaii Telescope (CFHT, on the Big Island of Hawaii) and the T\'elescope Bernard Lyot (TBL, at Pic du Midi Observatory in southern France).

\begin{table}
{\scriptsize
\begin{tabular}{llllr}
\hline \hline \noalign{\smallskip}
\noalign{\smallskip} \hline \noalign{\smallskip}
Star & Other  & Spectral   &$ T_{\rm eff}  $ & $v \sin i$   \\
     &   Name &   Type         &(K)          &(km/s)      \\
	\hline 
\noalign{\smallskip}
\multicolumn{5}{c}{\bf $\beta$ Cephei stars} \\
\noalign{\smallskip}
HD 886   & $\gamma$ Peg & B2 IV & $22500$ & $10 \pm 5 $ \\
HD 16582 & $\delta$ Cet & B2 IV & $23000$ & $ 8 \pm 5 $ \\
HD 29248 & $\nu$ Eri & B2 III & $23000$ & $34 \pm 5$  \\ 
HD 44743 & $\beta$ CMa & B1 II-II & $26000$ & $25 \pm 5$ \\
HD 46328 & $\xi^{1}$ Cma & B1 III & $27000$ & $14 \pm 5$  \\
HD 129929 & V836 Cen & B2        & $24000$  & $25 \pm 5$ \\ 
HD 207330 & 81 Cyg & B3 III & $18000$ &  $50 \pm 5 $    \\
HD 218376 & 1 Cas & B0.5 IV & $27000$ &    $ 25\pm 5 $   \\
\noalign{\smallskip}
\multicolumn{5}{c}{\bf Slowly-pulsating B stars} \\
\noalign{\smallskip}
HD 3379 & 53 Psc & B2.5 IV & $17300$   & $37 \pm 10$  	\\ 
HD 24587 & 33 Eri & B5 V, SB1 &$13900$ & $ 25 \pm 5 $ 	\\ 
HD 26326 & GU Eri & B5 IV & $15000$    & $19 \pm 10$ \\
HD 28114 & V1143 Tau & B6 IV & $14600$ & $ 14 \pm 5 $ 	\\
HD 34798 & YZ Lep & B5 IV-V & $15600$ & $ 41 \pm 10 $ 	 \\
HD 45284 & BD-07 1424 & B8, SB2 & $14500$ & $ 97 \pm 10$  \\
HD 46005 & V727 Mon & B8 V & $21000$ & $ 150 \pm 20 $  \\
HD 138764 & IU Lib & B6 IV & $14000$  & $32 \pm 5$     \\
HD 140873 & PT Ser & B8 III, SB2 & $13900$  & $88 \pm 10$ \\
HD 181558 & V4199 Sgr & B5 III  & $14700$ & $ 25 \pm 5 $ \\
HD 206540 & BD +10 4604 & B5 IV & $14000$ & $12 \pm 5$  \\	
HD 208057 & 16 Peg & B3 V, SB ? & $16700$ & $145 \pm 20$\\
\multicolumn{5}{c}{\bf B-type emission line stars} \\
\noalign{\smallskip}
HD 148184 & $\chi$ Oph & B2Vne     & $24000$ & $155 \pm 20  $  \\ 
HD 181615 & $\upsilon$ Sgr   & B2Vpe   & $23000$    & $33 \pm 5  $ \\
\noalign{\smallskip}
\hline		
\end{tabular}	
\caption[]{Stars observed in this investigation. Included are effective temperature (as reported by H06, H07, H09) and projected rotational velocity obtained as described in Sect. 3.  }
\label{starlist}	
}	
\end{table}

\begin{table*}
{\scriptsize
\begin{tabular}{lcrrrrrccrr}
\hline \hline \noalign{\smallskip}
\noalign{\smallskip} \hline \noalign{\smallskip}
HD &   JD   & Exp Time & Peak & LSD & LSD & Int Limits & Detection & Detection &$ B_\ell$  &  H06, H07, H09  \\
     & -2 400 000 & (sec)&S/N & S/N&   Gain          & km/s     &  Flag  & Probability&(G)    & $ B_\ell$ (G) \\
	\hline 
\noalign{\smallskip}
\multicolumn{10}{c}{\bf $\beta$ Cephei stars} \\
\noalign{\smallskip}
HD 886    &  $54817.7174$& 80  &  967&  10235 & 10.6 & -47 / 50& ND   &      0.001 &   $ 6 \pm 11$            &               \\
HD 16582 &  $54817.8043$ &  280   & 1021& 10808  & 10.6 &  -45 / 55 &  ND  &  0.001     &   $2 \pm 10$    &  $ {\bf -49 \pm 13 }$ \\ 
HD 29248 & $54489.98272$ & 200 & 503 & 4790 & 9.5 & -34 / 130 & ND & 0.005& $ 9 \pm 38 $ & $ -41 \pm 28 $ \\ 
HD 44743  & $54488.9702$ & 60 & 1015 & 11568& 11.4 & -34 / 100 & ND & 0.899& $ -31 \pm 13 $ &$ -44 \pm 29$\\
HD 46328 &  $54488.98129$ & 247 & 738 & 6230& 8.4 & -11 / 60 &{\bf DD} & 1.000 & ${\bf 338 \pm 11} $ & ${\bf 330 \pm 45 }$ \\
HD 129929 &  $54164.0772$ & 1600& 291 & 2886&  9.9 & -3 / 115 & ND & 0.026 & $ -14 \pm 35$ & $ -80 \pm 35 $\\ 
HD 207330  &  $54819.7901$ &320 &  757 & 8369  & 11.1  & -91 / 65 & ND   &  0.004     &      $8 \pm 24 $         &               \\ 
HD 218376  &  $54815.7721$ & 480 & 703 & 6678  & 9.5 & -89 / 76&  ND  &   0.001    &     $13 \pm 23$          &               \\ 
\noalign{\smallskip}
\multicolumn{10}{c}{\bf Slowly-pulsating B stars} \\
\noalign{\smallskip}
HD 3379 &$54376.9734$  & 600 &  660 & 10546 & 16.0 &   -138 / 78 & ND & 0.127& $ -97 \pm 77$ & $ {\bf 155 \pm 42} $ \\ 
        &$54352.6918$  & 2880 & 901 & 14374 & 16.0  & 49 / 81 & ND & 0.194 & $  -49 \pm 53  $         &  \\
        &$54353.5664$  & 2200 & 781 & 11785 & 15.1  & -115 / 76 &ND & 0.037 &  $ 1 \pm 40    $       &  \\
        &$54354.6115$  & 2400& 757 & 11869 &  15.7 & 54 / 87&ND  & 0.012 &  $ -68 \pm 77  $         &  \\
        &$54355.6268$  & 2800 & 834 & 12861 &  15.4 & 56 / 89   & ND & 0.005 &  $  81 \pm 64  $        &  \\
       &$54356.6592$  & 4520 & 877 & 13555 & 15.5   & -115 / 82 &ND & 0.346 &  $  46 \pm 35  $        &  \\
HD 24587 &$54373.9702$  & 300 & 745 & 12553& 16.8 & -19 / 79 &ND  & 0.154&$ -48 \pm 34$ & $ {\bf -353 \pm 82} $\\ 
HD 26326 &  $54374.9800$    & 600 & 588 & 12454& 21.2& -34 / 63 & ND & 0.066& $ -18 \pm 22 $ & $ -30 \pm 33 $ \\
HD 28114 & $54373.9793$  & 600 & 531 & 6831&  12.9 & -27 / 50 & ND & 0.632& $ -28 \pm 30 $ & $ {\bf 107 \pm 33} $\\
HD 34798 & $54375.9794$  & 700& 700  &7529& 10.8 & -40 / 71 & ND & 0.377& $ -10 \pm 35 $ & $ -99 \pm 45 $ \\
HD 45284 & $54488.9926$ & 1120 & 374 & 6192& 16.6 & -63 / 100 & ND & 0.642& $  -1 \pm 89 $ & $ -55 \pm 50$ \\
HD 46005 &$54489.9998$ & 1600& 380 & 5004& 13.2& 121 / 154 & ND & 0.676 &$6 \pm 203 $ & $2 \pm 79 $ \\
HD 138764 & $54168.0624$ & 600 & 897 & 15056& 16.8& -47 / 42 & ND & 0.048& $ -2 \pm 14$ & $ 146 \pm 57 $\\
HD 140873 & $54168.0728$ & 600 & 820 &13978& 17.0 & -89 / 86 & ND & 0.440 & $ 14 \pm 29$ & ${\bf 99 \pm 31 } $ \\
          &$54111.7501$ & 2400 & 611 &10203& 16.7& -102 / 76& ND & 0.024 & $ 32 \pm 67$            &   \\
          &$54111.7723$ & 2400 & 651 &11039& 17.0& -102 / 79 & ND & 0.052 & $ -35 \pm 63$             &   \\
          &$54114.7410$ & 1680 & 697 & 8725& 12.5& -109 / 65 & ND & 0.010 &$  19 \pm 74$              &   \\
          &$54116.7550$ & 1680 & 402 & 5866& 14.6&  -107 / 65 & ND & 0.114 &$  60 \pm 46$    &   \\
          &$54277.3901$ & 1680 & 764 & 12898& 16.9& -86 / 81& ND & 0.140 &$  26 \pm 50$          &   \\
          &$54277.4119$ & 1680 & 719 & 14012 & 19.5& -86 / 81& ND & 0.001 & $  15 \pm 44$          &   \\
          &$54278.3945$ & 1680 & 831 & 12985& 15.6&-81 / 92 & ND & 0.085 &  $  -19 \pm 50 $          &   \\
HD 181558  & $54293.5141$ & 2400 & 603& 10111 & 16.8 &-73 / 11& ND & 0.111 & $-25 \pm 21$ & ${\bf -104 \pm 32  }$ \\
          & $54293.5442$ & 2400 & 554 & 9324 & 16.8 & -71 / 11& ND & 0.299 &$-13 \pm 22 $&   \\
HD 206540 & $54374.9051$ & 600 & 524 & 8751& 16.7& -34 / 29 & ND & 0.041& $ -13 \pm 18 $ & $ -2 \pm 27 $ \\	
HD 208057 & $54371.9016$ & 600 & 1092 & 17295& 15.8 & -160 / 107& {\bf MD } & 0.998&$ -47 \pm 50 $& ${\bf -156 \pm 31 }$\\
\noalign{\smallskip}
\multicolumn{10}{c}{\bf B-type emission line stars} \\
\noalign{\smallskip}
HD 148184  & $54635.4480$ & 3600 & 1770  & 21815& 12.3 &204 / 258 &  ND     & 0.077   &    $ -73 \pm 63$                &${\bf 136 \pm 16}$ \\
           & $54636.4437$ & 3600 & 1435& 18051 & 12.6 & 217 / 245 &  ND     & 0.421  &     $13 \pm 75$                & \\
           & $54637.4400$ & 3600 &       &15025& & 217 / 245&  ND     & 0.617   &     $-96 \pm 112$                & \\
           & $54638.4821$ & 3600 & 1017 & 12637& 12.4& 207 / 234&  ND     & 0.021   &    $43 \pm 72$                 & \\
HD 181615  &  $54635.5881$ & 3600  &1786 & 22815& 12.7 & -55 / 45& ND     & 0.798       &   $  16 \pm 7 $               &${\bf 38 \pm 10}$ \\
           & $54636.5875$ & 3600 & 1594 & 20298 & 12.7 & -52 / 47 &  ND     & 0.005    & $-6 \pm 8$                  & \\
           & $54638.6214$ & 3600 & 1071 & 13386& 12.5 &-47 / 52 &  ND     & 0.201   & $-15 \pm 12$                 & \\
\noalign{\smallskip}
\hline		
\end{tabular}	
\caption[]{Log of observations and magnetic field measurements obtained with ESPaDOnS and NARVAL. The Julian Date of each observation is reported, along with the exposure time, peak signal-to-noise ratio (S/N) per 1.8~km/s pixel in the reduced spectrum, resultant S/N per 2.6~km/s LSD profile pixel and computed gain, Integration limits for evaluation of the longitudinal magnetic field, Stokes $V$ detection diagnosis (DD=definite detection, MD=marginal detection, ND=no detection) and probability, and the longitudinal field measurement obtained by evaluation of Eq. (1). Marginal and definite Stokes $V$ detections, and longitudinal field measurements significant at greater than $3\sigma$, are shown in bold. Also shown are the longitudinal fields reported by H06, H07 and H09.}
\label{samplefields}	
}	
\end{table*}



 \begin{figure}
  \centering
  \includegraphics[width=6.5cm, angle=-90]{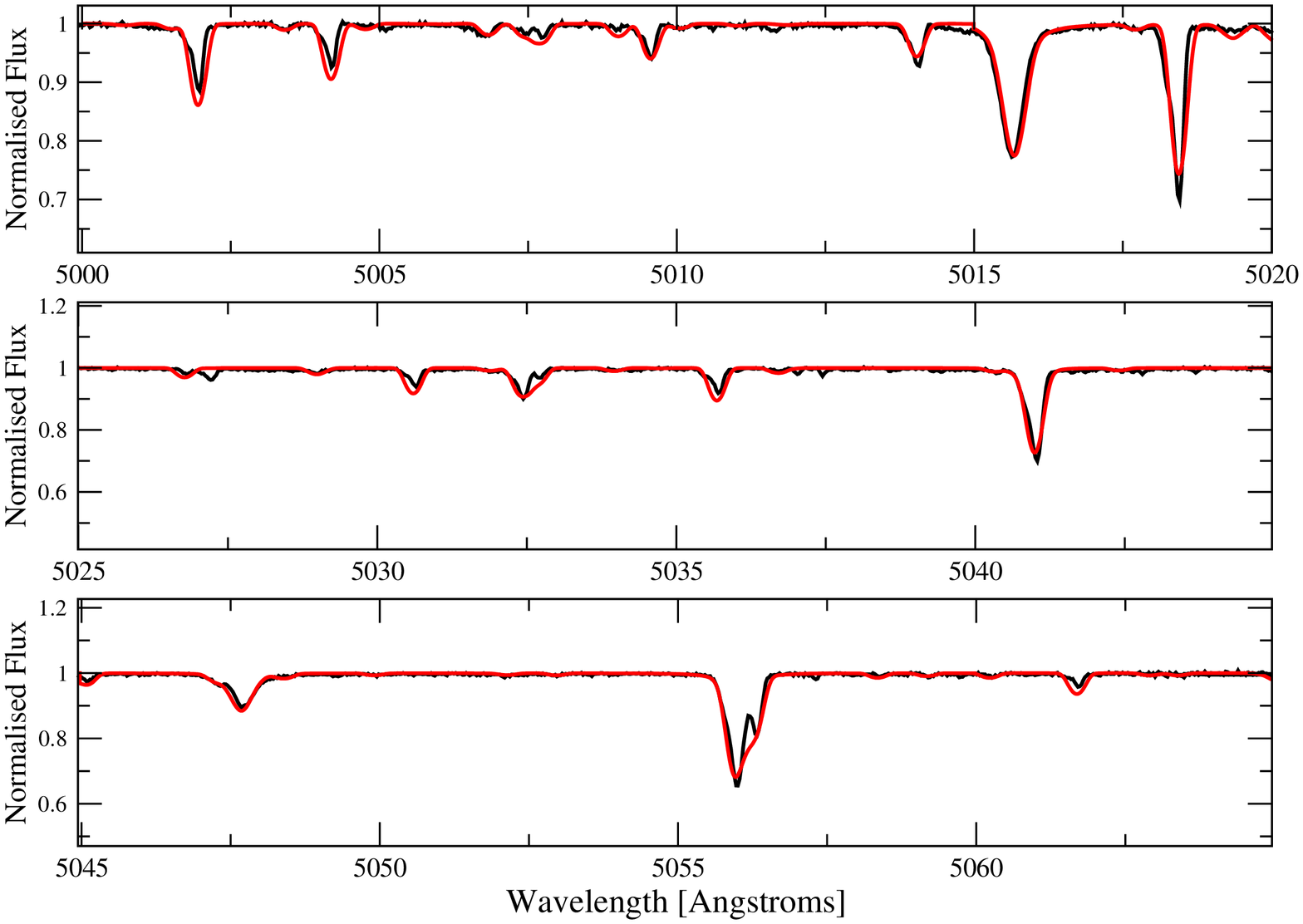}\\
    \includegraphics[width=6.5cm, angle=-90]{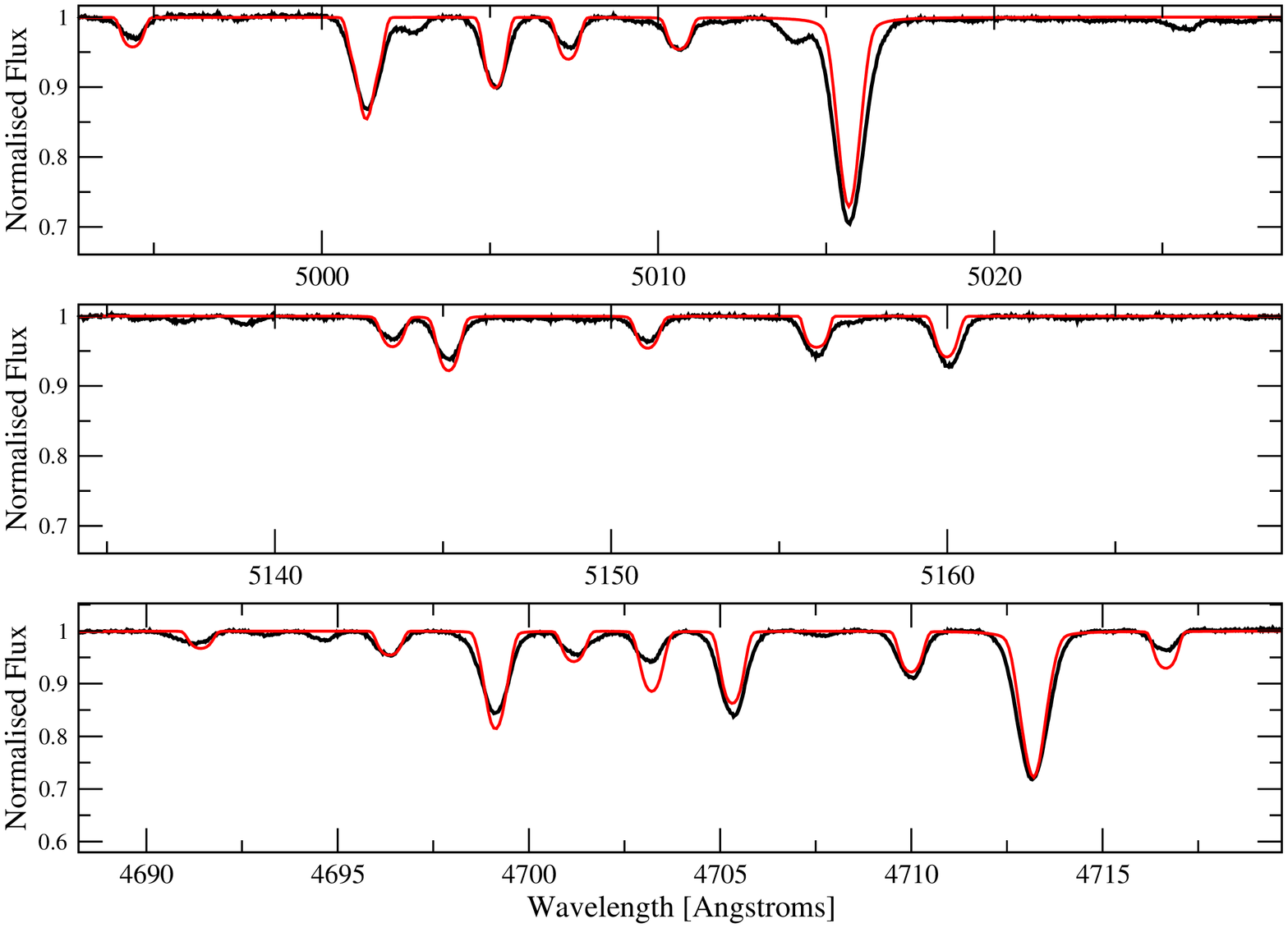}
  \caption{{\em Top frame --}\ Comparison between a solar abundance model spectrum (smooth red curve) for a temperature of 14000 K and the observation of HD 206540 (black crosses). {\em Lower frame --}\  Comparison between a solar abundance model spectrum (smooth red curve) for a temperature of 27000 K and the observation of HD 44743 (black crosses). The synthetic models acceptably reproduce the spectrum. This supports the use of solar abundance line masks in the LSD extraction.}
  \label{compare}
  \end{figure}

  \begin{figure*}
  \centering
  \includegraphics[width=0.99\textwidth]{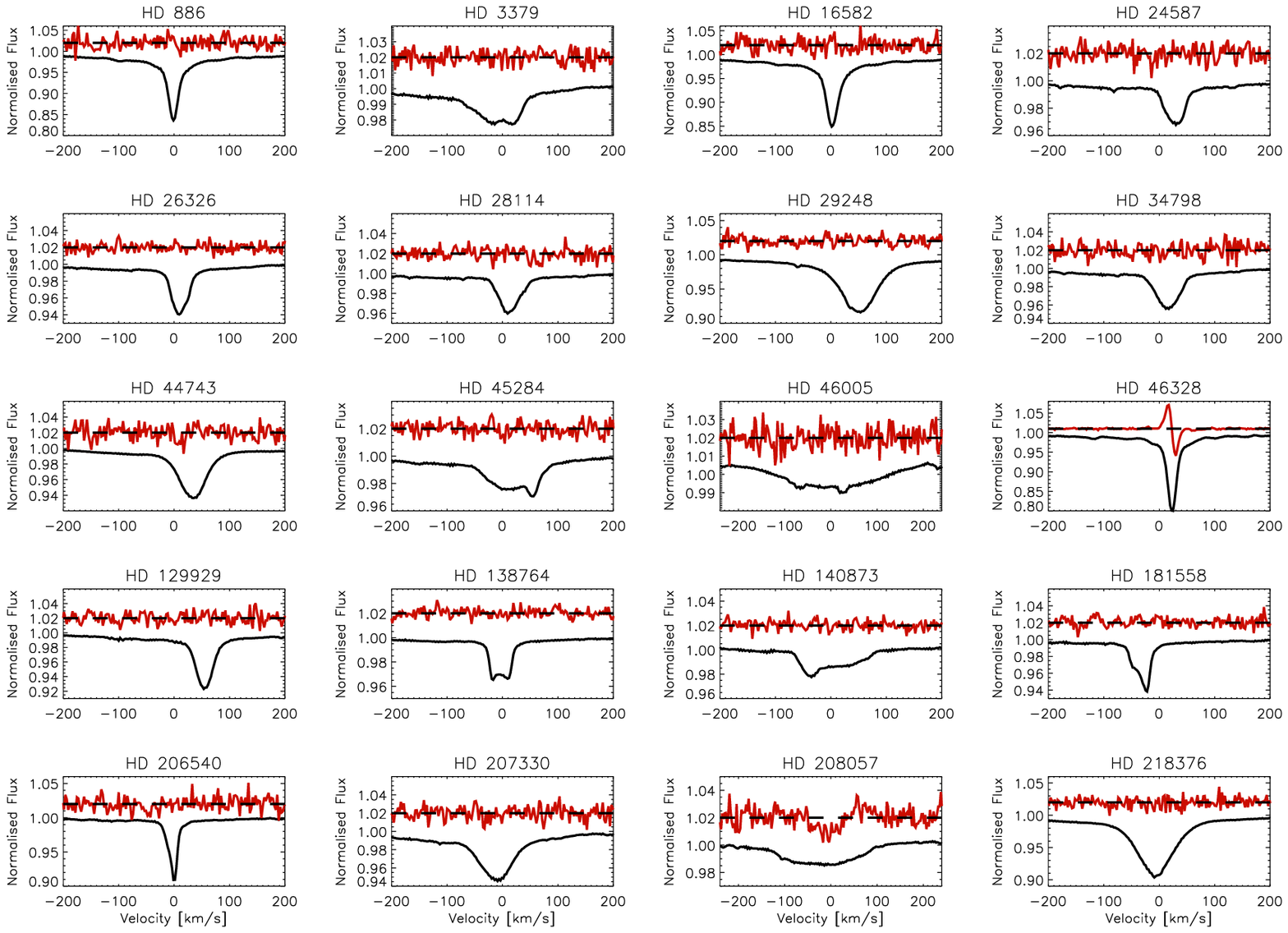}
  \caption{Least-Squares Deconvolved profiles of the SPB and $\beta$~Cep stars (for stars with multiple observations, a single example is shown). Each panel represents a different star, with the bottom curve showing the LSD mean intensity (Stokes $I$) profile and the top curve showing the circular polarisation Stokes $V$ profile. Note the frequently asymmetric or otherwise distorted profile shapes due to pulsation and binarity.} 
  \label{lsdplots}
  \end{figure*}

\section{Targets and observations}

Targets for this study were selected primarily from those stars observed by H09, and to a lesser extent by H06 and H07, that were observable from the CFHT and/or from TBL. Three additional $\beta$ Cep stars were also observed.  Because many of the stars observed by H06/H07/H09 are located in the southern hemisphere, only 19 of their targets could be observed from CFHT and Pic du Midi. The complete list of our 22 targets is provided in Table 1.

The selected sample of 8 $\beta$ Cep stars is shown in Table 1. H09 reported the detection of a $366\pm 11$~G longitudinal magnetic field in $\xi^{1}$ CMa and a field of $-49 \pm 13$~G in $\delta$ Cet. Four additional $\beta$~Cep stars were undetected by those authors, with typical uncertainties of about 30~G. The 3 remaining $\beta$~Cep stars HD 886, HD 207330 and HD 218376 were not observed by H06/07/09.

The SPB stars are of particular interest, because H09 reported that nearly half of their SPB sample is magnetic. Of the 12 SPB stars in our sample (all of which were observed by H06/H09), H06 and H09 claim detection of magnetic field in 6. Finally, as part of this study 2 {\bf B-type emission line stars} ($\upsilon$ Sgr\footnote{\cite{koubsky:2006} report that $\upsilon$~Sgr is in fact a mass-transfer binary in the initial phases of mass transfer, in which the emission lines arise from the disc.}  and $\chi$ Oph) were also observed. Both of the stars are considered by H07 to host weak magnetic fields. 

To evaluate the presence of magnetic fields in our targets, we have acquired high-resolution circular polarisation (Stokes $V$) spectroscopy of our targets. The observations reported here were obtained over a period of about a year, using the ESPaDOnS and NARVAL spectropolarimeters, the cross-dispersed echelle spectropolarimeters built for the Canada-France-Hawaii Telescope and Bernard Lyot Telescope, respectively. These instruments are effectively identical, and are conceptually similar to the MuSiCoS spectropolarimeter which was used extensively for high-precision magnetic measurements (e.g. \cite{wade:2000a}, \cite{donati:2001}, \cite{shorlin:2002}). ESPaDOnS and NARVAL are however characterised by a factor of about 20 times higher efficiency. 

The polarisation analyser is located at the Cassegrain focus of the telescope. Light passing though the pinhole aperture traverses a rotatable $\lambda$/2 waveplate, a fixed $\lambda$/4 waveplate, a second rotatable $\lambda$/2 waveplate, and finally a small-angle Wollaston prism, followed by a lens which refocuses the (now double) star image on the input of two optical fibres. This relatively complex polarisation analyser, which employs Fresnel rhombs, allows for essentially achromatic analysis over the large bandpass (3700~\AA\ to 1.04 $\mu$m). 

The two output beams from the Wollaston prism, which have been analysed into the two components of circular polarisation, are then carried by the pair of optical fibres to a stationary and temperature-controlled cross-dispersed spectrograph where two interleaved spectra are formed, covering virtually the entire desired wavelength range with a resolving power of $R\simeq 65 000$. The $I$ component of the stellar Stokes vector is formed by adding the two corresponding spectra, while the $V$ polarisation component is obtained essentially from the difference of the two spectra. To minimise systematic errors due to small misalignments, differences in transmission, effects of seeing, etc., one complete observation of a star consists of four successive spectra; for the second and third, the waveplate settings are changed so as to exchange the positions of the two analysed spectra on the CCD (c.f. \citealt{donati:1997}).  

The actual reduction of observations is carried out at the observatories using the dedicated software package Libre-ESpRIT, which yields both the $I$ spectrum and the $V$ circular polarisation spectrum of each star observed. Each reduced spectrum is normalised order-by-order using a FORTRAN code specifically optimised to fit the continuum of hot stars with emission lines.

A diagnostic null spectrum called the $N$ spectrum, computed by combining the four observations of polarisation in such a way as to have real polarisation cancel out, is also calculated by Libre-Esprit. The $N$ spectrum tests the system for spurious polarisation signals. In all of our observations, the $N$ spectrum is quite featureless, as expected. The final spectra consist of ascii files tabulating $I$, $V$, $N$, and the estimated uncertainty per pixel as a function of wavelength, order by order. 


In this study, 40 spectra of the 22 B-type targets were acquired. The peak signal-to-noise ratios of the reduced 1-dimensional spectra range from 300 to 1700 per 1.8 km/s spectral pixel. The log of observations is reported in Table 2.

\section{Analysis}

The spectra were analysed using the Least-Squares Deconvolution (LSD) multiline analysis method \citep{donati:1997}. LSD produces mean Stokes $I$ and $V$ profiles using essentially all metallic and He lines in the stellar spectrum, assuming that the observed spectrum can be represented as the convolution of a single "mean" line profile with an underlying spectrum of unbroadened metal and helium lines of appropriate wavelength, depth and Land\'e factor (the "line mask" computed using spectrum synthesis; c.f. \citealt{wade:2000a}). The LSD model allows the computation of single, average Stokes $I$ and $V$ line profiles, representative of essentially all lines in the stellar spectrum, usually characterised by a signal-to-noise ratio dramatically higher than that of individual spectral lines. Line masks for this study were compiled using Vienna Atomic Line Database (VALD, \citealt{kupka:1999}) "extract stellar" requests, with effective temperature and surface gravity appropriate to each target, and assuming solar abundances. As discussed by \cite{shorlin:2002}, the LSD S/N is only weakly sensitive to the line-depth cutoff employed to populate the mask. Following their results, we have chosen to employ a line-depth cutoff equal to 10\% of the continuum. Imposing such a cutoff has a related advantage: because weaker lines are less likely to have published experimental Land\'e factors (and to generally have more poorly-determined atomic data), we pre-filter our line list to (statistically) exclude those lines with the poorest data.  Except for the line-depth cutoff, elimination of Balmer lines, and restriction of the mask to the ESPaDOnS/Narval spectral range, no other line selection has generally been performed.

To verify the suitability of these masks, synthetic spectra corresponding to the mask parameters and content were computed and compared to the observations. For SPB and $\beta$~Cep stars, an acceptable agreement was always obtained. For Be stars, an additional filtering of the mask was performed to exclude lines identified in the observations as being significantly affected by circumstellar emission. The agreement between a solar abundance synthetic spectrum and the observation of SPB star HD 206540 and the $\beta$~Cep star HD 44743 for a small wavelength window is illustrated in Fig. 1. 
For a typical SPB star ($T_{\rm eff}$ = 14000 K), 520 lines were included in the mask. For a typical $\beta$ Cep star ($T_{\rm eff}$ = 25000 K), 260 were included. For the two B-type emission line stars the filtered mask contained 177 lines. Although the number of strong lines, as well as the total number of lines, decreases with increasing temperature, the fraction of lines between a depth of 0.1 and 0.3 is around 80 percent for all line masks used. 

 \begin{figure}
  \centering
  \includegraphics[width=8.5cm]{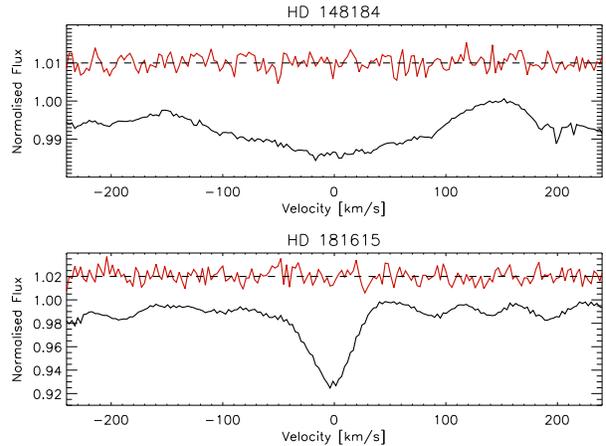}
  \caption{Representative Least-Squares Deconvolved profiles the two {\bf B-type emission line} stars. Each panel represents a different star, with the bottom curve showing the LSD mean intensity (Stokes $I$) profile and the top curve showing the circular polarisation Stokes $V$ profile. Note the relatively poor quality of the deconvolution (distortion of the continuum, for example, due to the smaller number of lines in the mask) as compared to the profiles shown in Fig. 2.} 
  \label{lsdplotsbe}
  \end{figure}

 \begin{figure}
  \centering
  \includegraphics[angle=-90,width=9cm]{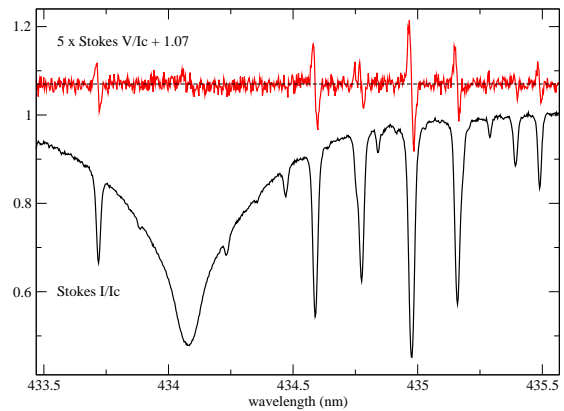}
  \caption{Stokes $V$ signatures in individual lines in the spectrum of $\xi^1$ CMa in the region of H$\gamma$.}
  \label{xicma}
  \end{figure}

The LSD procedure propagates the formal error bars associated with each spectral pixel through the deconvolution. Because the computed LSD spectrum (the convolution of the deconvolved mean profile with the line mask) represents a highly simplified model of the spectrum, the overall agreement between the model and observations is generally relatively poor. To bring the observed and model spectrum into formal agreement, we interpret their mutual differences as random noise and scale the error bars associated with each pixel in the LSD profile in order to yield a final reduced $\chi^2$ of the model relative to the observations that is equal to unity \citep{wade:2000a}. This has the effect of assigning errors to the LSD profile that are reflective of the real uncertainties of the modeling. When a magnetic field is present, each spectral line produces an associated Stokes $V$ signature, resulting in a relatively complex Stokes $V$ spectrum. Because the LSD model is unable to reproduce the details of this spectrum, the discrepancies are relatively severe, significant scaling of the error bars is required, and the uncertainties are dominated by the limitations of the model. On the other hand, when no magnetic field is present, the Stokes $V$ spectrum is featureless. The model is able to fit the Stokes $V$ spectrum essentially within the formal errors, relatively little scaling is required, and the uncertainties are dominated by photon noise. The procedures used to derive the uncertainties have been extensively tested, and shown to generate assigned uncertainties which are consistent with the real scatter of observations for both magnetic and non-magnetic standards (e.g. \citealt{wade:2000a}, \citealt{shorlin:2002}, \citealt{chadid:2004}, \citealt{alecian:2008b}). The LSD profiles computed from the new spectra, one for each of the sample stars, are presented in Figs. 2 and 3.

The presence or absence of a magnetic field detection was evaluated using two methods. First, we used the statistical test described by \cite*{donati:1992} and \cite{donati:1997} to diagnose the presence of a signal in the LSD $V$ and $N$ profiles. In this method, a signal is "definitely" detected if the associated detection probability of Stokes $V$ within the spectral line is larger than 99.999 per cent (corresponding to a false alarm probability smaller than $10^{-5}$), and if the detection probabilities both outside the line and in the diagnostic null within the line are insignificant. A "marginal" detection corresponds to a detection probability between 99.9\% and 99.999\% (false alarm probability between $10^{-3}$ and $10^{-5}$). A detection probability below 99.9\% corresponds to no formal detection.

The second method involved inferring the mean longitudinal magnetic field, evaluated by computing the first-order moment of the Stokes $V$ profile within the line according to:

\begin{equation}
\displaystyle 
B_\ell =  -2.14\times 10^{11}  \frac {\int vV(v) dv} {\lambda g c \int [1-I(v)] dv}
\end{equation}

\noindent  ( \citealt{mathys:1989}, \citealt{wade:2000a}) where $g$ is the mean Land\'e factor and $\lambda$  is the mean wavelength of all the lines included in the mask. LSD profiles were locally re-normalised to a continuum level of 1.0 before evaluation of Eq. (1). Uncertainties associated with $B_\ell$ were computed by propagating the formal uncertainties of each LSD spectral pixel through Eq. (1).

The integration range (in km/s) associated with Eq. (1) was computed individually for each LSD profile. First, we identified the first and last point in the Stokes $I$ profile for which the flux was equal to 85\% of the continuum. To correct for errors due to line profile structure (e.g. binarity), these preliminary integration ranges were then visually inspected and adjusted to best match the observed profile span. The integration ranges employed for each profile are reported in Table 2. While this procedure can result in integration ranges of different LSD profiles of the same star differing significantly (for example variable the SB2 HD 140873 in Table 2), this careful adjustment of the integration range allows us to optimize the magnetic diagnosis to the detailed shape of the line profile. This therefore represents an important strength of this technique - one that is not possible at the low resolution of FORS1.

It is important to note that the longitudinal field is used only as a statistical indicator of the field strength and not as the primary diagnostic of the presence of a magnetic field. This is because a large variety of magnetic configurations can produce a mean longitudinal field component that is formally null (i.e. for which the first-order moment of Stokes $V$ is zero - a profile approximately symmetric about the centre-of-gravity of the line). However, nearly all of these configurations will generate a detectable Stokes $V$ signature in the velocity-resolved line profile. As a consequence, high-resolution spectropolarimetry combined with LSD has been repeatedly shown to be effective at detecting relatively complex magnetic configurations, in both hot (e.g. \citealt{donati:2006}) and cool (e.g. \citealt{donati:1999}) stars.

In addition to measuring the longitudinal magnetic field from each LSD profile, we also inferred projected rotational velocities using the automated fitting procedure included in the BinMag IDL visualisation tool (constructed by O. Kochukhov). This procedure compares individual observed line profiles with synthesized profiles computed for appropriate temperature and gravity. Multiple metallic lines were fit in each spectrum to achieve a distribution of $v\sin i$ measurements. The resulting means and deviations are reported in Table 1. The derived uncertainties are relatively high due to the distortion of line profiles by pulsation and binarity. The majority of the derived $v\sin i$s are in good agreement with those reported by H06, although our adopted uncertainties are somewhat larger.

Our measured detection probabilities and longitudinal fields, as well as the FORS1 longitudinal fields reported by H06, H07 and H09, are reported in Table 2.

\section{Results}

In our sample of 22 B-type stars, we obtain 1 definite detection, one marginal detection, and 20 non-detections of Stokes $V$ signatures reflecting the presence of photospheric magnetic fields. Among the $\beta$ Cep stars, we obtain a clear detection of magnetic field in $\xi^{1}$ CMa, which shows strong Stokes $V$ signatures in the LSD profile (Fig. 2), as well as in individual spectral lines (Fig. \ref{xicma}). The longitudinal magnetic field measured from our single spectrum of this star is $338\pm 11$~G, formally consistent with the measurement by H06 and H09. No fields are detected in the remaining 7 $\beta$ Cep stars, with typical longitudinal field formal errors of a few tens of G. In contrast, H09 report the $\beta$~Cep star $\delta$ Cet to be magnetic ($B_\ell=-49 \pm 13 $~G), a result that we do not confirm ( $B_\ell=2\pm 10$~G).

Among the SPB stars,  we obtain a marginal detection of a Stokes $V$ signature in the LSD profile of 16 Peg. Additional NARVAL observations of this star \citep{henrichs:2009} confirm the detection of the field, and provide constraints on the field geometry and the stellar rotational period. None of the remaining SPB targets shows any evidence of magnetic field, with a median longitudinal field error bar of 45 G. Of SPB stars which were not detected by us, H09 claim detections of 5: HD 3379, HD 24587, HD 28114 , HD 140873, HD 181588. We find no evidence of fields in these stars, with error bars comparable to those of H06 and H09. Moreover, for three of these stars we have obtained multiple null results. Given the high quality of the spectra and the sensitivity of high-resolution spectropolarimetry to the magnetic field at essentially all rotational phases, we conclude that it is highly likely that none of these 5 SPB stars is in fact magnetic.

Finally, in the {\bf two B-type emission line stars} observed in this study (HD 148184 and HD 181615) no magnetic fields are detected. Multiple observations were acquired for both stars. Longitudinal field error bars smaller than 10~G were achieved for HD 181615, allowing us to conclude confidently that no field is present. However, the error bars for the broader-lined star HD 148184 are rather larger, and while we find no evidence of a field at the $\sim 150$~G level as reported by H07, such a field could in principle remain undetected in our observations.


\section{Discussion and Conclusions}

The basic results of this study are the confirmation of magnetic fields in the $\beta$~Cep star $\xi^1$~CMa (HD 46328) and the SPB star 16~Peg (HD 208057), and the failure to detect fields in 20 other $\beta$~Cep, SPB and Be stars, 8 of which were claimed to be magnetic by H06, H07 and H09. The $\beta$~Cep star $\xi^1$~CMa has relatively sharp lines ($v\sin i=14$~km/s), and we infer a strong magnetic field ($B_\ell = 338 \pm 11$~G) which is consistent with that reported by H09. 
The SPB star 16 Peg is a relatively rapid rotator, with $v\sin i=145$~km/s. We measure a weak longitudinal magnetic field of $47\pm 50$~G\footnote {Recall that although the longitudinal field is not significantly detected, circular polarisation is unambiguously detected in the LSD profile.}. \cite{henrichs:2009} have obtained many NARVAL Stokes $V$ spectra of this star in which the Stokes $V$ signature was detected repeatedly. Those authors inferred an organised surface magnetic field with longitudinal field extrema of $\approx$ $\pm 140$ G.


For the other 8 stars reported to be magnetic by H06/07/09, we are unable to confirm their detections. While the generation of spurious field {\em detections} due to instrumental and astrophysical effects is a real danger in stellar magnetometry, it is very difficult to engineer scenarios that produce spurious {\em non-detections} of Stokes $V$ signatures, particularly in a systematic manner for a large sample of stars. It therefore seems very unlikely that the null results obtained here for many stars are a result of a systematic inability to detect magnetic fields. This is supported by our detection of magnetic fields in 16 Peg and $\xi^1$~CMa (notwithstanding the fact that the longitudinal field of 16 Peg is formally null). That our error bars are not overestimated is supported by the reduced $\chi^2$ of the longitudinal fields we measure (see e.g. \citealt{bohlender:1993}): For the non-detections, the longitudinal field reduced $\chi^2$ is 0.96, indicating that the longitudinal field error bars are fully consistent with the scatter of the measurements. In addition, the distribution of longitudinal fields measured from the diagnostic $N$ profiles (which should correspond to zero magnetic field) are statistically consistent with the expected normal distribution about a mean of zero, with a dispersion in agreement with the inferred Stokes $V$ error bars.

The Stokes $V$ signature amplitude will vary due to stellar rotation: could it be that we have observed these stars when the signature is weak and undetectable? While possible, this is statistically unlikely. First, if we take the results of H06/07/09 at face value, then the magnetic strengths and geometries of these stars must favour magnetic detection (because so many were detected by them). Secondly, it can be easily confirmed by inspecting the Stokes $V$ profile phase variations of real magnetic stars (e.g. \citealt{wade:2000b}) that the signature amplitude remains relatively constant as a function of stellar rotation. Both of these facts suggest that our failure to detect magnetic fields in these stars is unrelated to phase or geometry effects.

An examination of Fig. 2 reveals that many of the LSD profiles show asymmetries and distortions (HD 3379, HD 45284, HD 46005, HD 138764, HD 140873, HD 181558, HD 208056). Could it be that these properties impair our ability to detect Stokes $V$ signatures? The peculiar line profile shapes likely result from a variety of processes, in particular binarity and pulsation. For example, both HD 45284 and HD 140873 are both confirmed SB2s. The other stars showing significant profile asymmetries may well also be spectroscopic binaries. However, the presence of the line profile of a companion to a magnetic star will not reduce or otherwise modify the Stokes $V$ signature observed in high-resolution spectropolarimetry. Therefore, although the inferred longitudinal field and its uncertainty maybe modified somewhat, (because we infer the field by integration over the entire composite profile) our ability to {\em detect} fields is not substantially affected.

Naturally, as this study focuses on pulsating stars, many of the observed profiles are distorted by pulsation. HD 138764 and 181558 appear to represent extreme examples of this phenomenon. While pulsation velocity fields will distort Stokes $V$ profiles in a manner analogous to Stokes $I$ (see \cite{donati:2001} for a nice illustration of this effect), there is no evidence to suggest that the Stokes $V$ profiles are significantly more difficult to detect. In fact, the rapid temporal changes in line profile shapes due to pulsation are more likely to introduce {spurious} Stokes $V$ {\em detections \citep{schnerr:2006}}, rather than to cause Stokes $V$ signatures to disappear.  It is also important to appreciate that the effects of pulsation and binarity, while discussed here in the context of high-resolution spectropolarimetry, are equally significant for low-resolution spectropolarimetry. The major difference is that high-resolution spectra provide us with the capability to see and diagnose these effects, whereas they are essentially invisible in low-resolution spectra.


In conclusion: the present study confirms fields in only two stars of the H06/07/09 sample: $\xi^1$ CMa and 16 Peg. We do not confirm the remaining detections. This inability to confirm the reported detections made using FORS1 poses a real problem, and clearly illustrates the need for further investigation. To underscore this need, we point out that of the 10 stars in our sample reported to be magnetic by H06/07/09, just 3 are claimed by them to be detected at more than 4.5$\sigma$ confidence - and we have confirmed fields in two of those. None of the remaining stars is detected by them at high significance. In fact, if their error bars were underestimated by just one-third ($\sim 33$\%, the difference between $\sigma=30$~G and $\sigma=40$~G), most of those detections would be transformed into sub-$3\sigma$ non-detections. To our knowledge, no detailed investigation of the accuracy of the FORS1 error bars derived by these authors has ever been performed (in particular error bars below a few tens of G).

Outside of the present study and those of H06/07/09, there are only a few cases of confirmed magnetic fields in $\beta$~Cephei and SPB stars ($\beta$ Cep itself, \citep{henrichs:2000a}, V2052 Oph \citep{neiner:2003a}, $\zeta$ Cas \citep{neiner:2003b}), and no confirmed detections of magnetic fields in classical Be stars. Magnetic fields therefore appear to be relatively rare in these classes of stars. We therefore propose that the incidence of fields with strengths detectable with current spectropolarimeters in these classes of stars merely reflects the (apparently rather low) incidence of fields in the larger population of stars at these spectral types, and is fundamentally unrelated to the physical properties that define these particular classes of stars.  


\bsp

\label{lastpage}

\bibliographystyle{mn2e}
\bibliography{references}

\begin{thebibliography}{}

\bibitem[\protect\citeauthoryear{{Alecian}, {Catala}, {Wade}, {Donati},
  {Petit}, {Landstreet}, {B{\"o}hm}, {Bouret}, {Bagnulo}, {Folsom}, {Grunhut}
  \& {Silvester}}{{Alecian} et~al.}{2008}]{alecian:2008b}
{Alecian} E.,  {Catala} C.,  {Wade} G.~A.,  {Donati} J.-F.,  {Petit} P.,
  {Landstreet} J.~D.,  {B{\"o}hm} T.,  {Bouret} J.-C.,  {Bagnulo} S.,  {Folsom}
  C.,  {Grunhut} J.,    {Silvester} J.,  2008, \mnras, 385, 391

\bibitem[\protect\citeauthoryear{{Alecian}, {Wade}, {Catala}, {Bagnulo},
  {Boehm}, {Bohlender}, {Bouret}, {Donati}, {Folsom}, {Grunhut} \&
  {Landstreet}}{{Alecian} et~al.}{2008}]{alecian:2008a}
{Alecian} E.,  {Wade} G.~A.,  {Catala} C.,  {Bagnulo} S.,  {Boehm} T.,
  {Bohlender} D.,  {Bouret} J.-C.,  {Donati} J.-F.,  {Folsom} C.~P.,  {Grunhut}
  J.,    {Landstreet} J.~D.,  2008, \aap, 481, L99

\bibitem[\protect\citeauthoryear{{Bohlender}, {Landstreet}, {Brown} \&
  {Thompson}}{{Bohlender} et~al.}{1987}]{bohlender:1987}
{Bohlender} D.~A.,  {Landstreet} J.~D.,  {Brown} D.~N.,    {Thompson} I.~B.,
  1987, \apj, 323, 325

\bibitem[\protect\citeauthoryear{{Bohlender}, {Landstreet} \&
  {Thompson}}{{Bohlender} et~al.}{1993}]{bohlender:1993}
{Bohlender} D.~A.,  {Landstreet} J.~D.,    {Thompson} I.~B.,  1993, \aap, 269,
  355

\bibitem[\protect\citeauthoryear{{Borra} \& {Landstreet}}{{Borra} \&
  {Landstreet}}{1979}]{borra:1979}
{Borra} E.~F.,  {Landstreet} J.~D.,  1979, \apj, 228, 809

\bibitem[\protect\citeauthoryear{{Borra}, {Landstreet} \& {Thompson}}{{Borra}
  et~al.}{1983}]{borra:1983}
{Borra} E.~F.,  {Landstreet} J.~D.,    {Thompson} I.,  1983, \apjs, 53, 151

\bibitem[\protect\citeauthoryear{{Chadid}, {Wade}, {Shorlin} \&
  {Landstreet}}{{Chadid} et~al.}{2004}]{chadid:2004}
{Chadid} M.,  {Wade} G.~A.,  {Shorlin} S.~L.~S.,    {Landstreet} J.~D.,  2004,
  \aap, 413, 1087

\bibitem[\protect\citeauthoryear{{Donati}, {Catala}, {Wade}, {Gallou},
  {Delaigue} \& {Rabou}}{{Donati} et~al.}{1999}]{donati:1999}
{Donati} J.-F.,  {Catala} C.,  {Wade} G.~A.,  {Gallou} G.,  {Delaigue} G.,
  {Rabou} P.,  1999, \aaps, 134, 149

\bibitem[\protect\citeauthoryear{{Donati}, {Howarth}, {Bouret}, {Petit},
  {Catala} \& {Landstreet}}{{Donati} et~al.}{2006}]{donati:2006}
{Donati} J.-F.,  {Howarth} I.~D.,  {Bouret} J.-C.,  {Petit} P.,  {Catala} C.,
   {Landstreet} J.,  2006, \mnras, 365, L6

\bibitem[\protect\citeauthoryear{{Donati}, {Semel}, {Carter}, {Rees} \&
  {Collier Cameron}}{{Donati} et~al.}{1997}]{donati:1997}
{Donati} J.-F.,  {Semel} M.,  {Carter} B.~D.,  {Rees} D.~E.,    {Collier
  Cameron} A.,  1997, \mnras, 291, 658

\bibitem[\protect\citeauthoryear{{Donati}, {Semel} \& {Rees}}{{Donati}
  et~al.}{1992}]{donati:1992}
{Donati} J.-F.,  {Semel} M.,    {Rees} D.~E.,  1992, \aap, 265, 669

\bibitem[\protect\citeauthoryear{{Donati}, {Wade}, {Babel}, {Henrichs}, {de
  Jong} \& {Harries}}{{Donati} et~al.}{2001}]{donati:2001}
{Donati} J.-F.,  {Wade} G.~A.,  {Babel} J.,  {Henrichs} H.~F.,  {de Jong}
  J.~A.,    {Harries} T.~J.,  2001, \mnras, 326, 1265

\bibitem[\protect\citeauthoryear{{Henrichs}, {de Jong}, {Donati}, {Catala},
  {Wade}, {Shorlin}, {Veen}, {Nichols} \& {Kaper}}{{Henrichs}
  et~al.}{2000}]{henrichs:2000a}
{Henrichs} H.~F.,  {de Jong} J.~A.,  {Donati} J.-F.,  {Catala} C.,  {Wade}
  G.~A.,  {Shorlin} S.~L.~S.,  {Veen} P.~M.,  {Nichols} J.~S.,    {Kaper} L.,
  2000, in {Smith} M.~A.,  {Henrichs} H.~F.,   {Fabregat} J.,  eds, ASP Conf.
  Ser. 214: IAU Colloq. 175: The Be Phenomenon in Early-Type Stars {The
  Magnetic Field of {$\beta$} Cep and the Be Phenomenon}.
p.~324

\bibitem[\protect\citeauthoryear{{Henrichs}, {Neiner}, {Schnerr}, {Verdugo},
  {Alecian}, {Catala}, {Cochard}, {Guti{\'e}rrez}, {Huat}, {Silvester} \&
  {Thizy}}{{Henrichs} et~al.}{2009}]{henrichs:2009}
{Henrichs} H.~F.,  {Neiner} C.,  {Schnerr} R.~S.,  {Verdugo} E.,  {Alecian} A.,
   {Catala} C.,  {Cochard} F.,  {Guti{\'e}rrez} J.,  {Huat} A.-L.,  {Silvester}
  J.,    {Thizy} O.,  2009, in IAU Symposium Vol.~259 of IAU Symposium, {The
  magnetic field of the B3V star 16 Pegasi}.
pp 393--394

\bibitem[\protect\citeauthoryear{{Hubrig}, {Briquet}, {De Cat}, {Sch{\"o}ller},
  {Morel} \& {Ilyin}}{{Hubrig} et~al.}{2009}]{hubrig:2009}
{Hubrig} S.,  {Briquet} M.,  {De Cat} P.,  {Sch{\"o}ller} M.,  {Morel} T.,
  {Ilyin} I.,  2009, Astronomische Nachrichten, 330, 317

\bibitem[\protect\citeauthoryear{{Hubrig}, {Briquet}, {Sch{\"o}ller}, {De Cat},
  {Mathys} \& {Aerts}}{{Hubrig} et~al.}{2006}]{hubrig:2006}
{Hubrig} S.,  {Briquet} M.,  {Sch{\"o}ller} M.,  {De Cat} P.,  {Mathys} G.,
  {Aerts} C.,  2006, \mnras, 369, L61

\bibitem[\protect\citeauthoryear{{Hubrig}, {Yudin}, {Pogodin}, {Sch{\"o}ller}
  \& {Peters}}{{Hubrig} et~al.}{2007}]{hubrig:2007}
{Hubrig} S.,  {Yudin} R.~V.,  {Pogodin} M.,  {Sch{\"o}ller} M.,    {Peters}
  G.~J.,  2007, Astronomische Nachrichten, 328, 1133

\bibitem[\protect\citeauthoryear{{Koubsk{\'y}}, {Harmanec}, {Yang},
  {Netolick{\'y}}, {{\v S}koda}, {{\v S}lechta} \& {Kor{\v
  c}{\'a}kov{\'a}}}{{Koubsk{\'y}} et~al.}{2006}]{koubsky:2006}
{Koubsk{\'y}} P.,  {Harmanec} P.,  {Yang} S.,  {Netolick{\'y}} M.,  {{\v
  S}koda} P.,  {{\v S}lechta} M.,    {Kor{\v c}{\'a}kov{\'a}} D.,  2006, \aap,
  459, 849

\bibitem[\protect\citeauthoryear{{Kupka}, {Piskunov}, {Ryabchikova}, {Stempels}
  \& {Weiss}}{{Kupka} et~al.}{1999}]{kupka:1999}
{Kupka} F.,  {Piskunov} N.,  {Ryabchikova} T.~A.,  {Stempels} H.~C.,    {Weiss}
  W.~W.,  1999, \aaps, 138, 119

\bibitem[\protect\citeauthoryear{{Mathys}}{{Mathys}}{1989}]{mathys:1989}
{Mathys} G.,  1989, Fundamentals of Cosmic Physics, 13, 143

\bibitem[\protect\citeauthoryear{{Neiner}, {Geers}, {Henrichs}, {Floquet},
  {Fr{\' e}mat}, {Hubert}, {Preuss} \& {Wiersema}}{{Neiner}
  et~al.}{2003}]{neiner:2003a}
{Neiner} C.,  {Geers} V.~C.,  {Henrichs} H.~F.,  {Floquet} M.,  {Fr{\' e}mat}
  Y.,  {Hubert} A.-M.,  {Preuss} O.,    {Wiersema} K.,  2003, \aap, 406, 1019

\bibitem[\protect\citeauthoryear{{Neiner}, {Henrichs}, {Floquet}, {Fr{\'
  e}mat}, {Preuss}, {Hubert}, {Geers}, {Tijani}, {Nichols} \&
  {Jankov}}{{Neiner} et~al.}{2003}]{neiner:2003c}
{Neiner} C.,  {Henrichs} H.~F.,  {Floquet} M.,  {Fr{\' e}mat} Y.,  {Preuss} O.,
   {Hubert} A.-M.,  {Geers} V.~C.,  {Tijani} A.~H.,  {Nichols} J.~S.,
  {Jankov} S.,  2003, \aap, 411, 565

\bibitem[\protect\citeauthoryear{{Neiner}, {Hubert}, {Fr{\' e}mat}, {Floquet},
  {Jankov}, {Preuss}, {Henrichs} \& {Zorec}}{{Neiner}
  et~al.}{2003}]{neiner:2003b}
{Neiner} C.,  {Hubert} A.-M.,  {Fr{\' e}mat} Y.,  {Floquet} M.,  {Jankov} S.,
  {Preuss} O.,  {Henrichs} H.~F.,    {Zorec} J.,  2003, \aap, 409, 275

\bibitem[\protect\citeauthoryear{{Petit}, {Wade}, {Drissen}, {Montmerle} \&
  {Alecian}}{{Petit} et~al.}{2008}]{petit:2008}
{Petit} V.,  {Wade} G.~A.,  {Drissen} L.,  {Montmerle} T.,    {Alecian} E.,
  2008, \mnras, 387, L23

\bibitem[\protect\citeauthoryear{{Schnerr}, {Verdugo}, {Henrichs} \&
  {Neiner}}{{Schnerr} et~al.}{2006}]{schnerr:2006}
{Schnerr} R.~S.,  {Verdugo} E.,  {Henrichs} H.~F.,    {Neiner} C.,  2006, \aap,
  452, 969

\bibitem[\protect\citeauthoryear{{Shorlin}, {Wade}, {Donati}, {Landstreet},
  {Petit}, {Sigut} \& {Strasser}}{{Shorlin} et~al.}{2002}]{shorlin:2002}
{Shorlin} S.~L.~S.,  {Wade} G.~A.,  {Donati} J.-F.,  {Landstreet} J.~D.,
  {Petit} P.,  {Sigut} T.~A.~A.,    {Strasser} S.,  2002, \aap, 392, 637

\bibitem[\protect\citeauthoryear{{Stankov} \& {Handler}}{{Stankov} \&
  {Handler}}{2005}]{stankov:2005}
{Stankov} A.,  {Handler} G.,  2005, \apjs, 158, 193

\bibitem[\protect\citeauthoryear{{Wade}, {Donati}, {Landstreet} \&
  {Shorlin}}{{Wade} et~al.}{2000a}]{wade:2000a}
{Wade} G.~A.,  {Donati} J.-F.,  {Landstreet} J.~D.,    {Shorlin} S.~L.~S.,
  2000a, \mnras, 313, 851

\bibitem[\protect\citeauthoryear{{Wade}, {Donati}, {Landstreet} \&
  {Shorlin}}{{Wade} et~al.}{2000b}]{wade:2000b}
{Wade} G.~A.,  {Donati} J.-F.,  {Landstreet} J.~D.,    {Shorlin} S.~L.~S.,
  2000b, \mnras, 313, 823

\end{thebibliography}

\end{document}